\documentclass[twocolumn,aps,prl,showpacs,floatfix,superscriptaddress]{revtex4}
\pdfoutput=1
\usepackage[latin9]{inputenc}
\usepackage{units}
\usepackage{amsmath}
\usepackage{amssymb}
\usepackage{graphicx}

\makeatletter
\@ifundefined{textcolor}{}
{%
 \definecolor{BLACK}{gray}{0}
 \definecolor{WHITE}{gray}{1}
 \definecolor{RED}{rgb}{1,0,0}
 \definecolor{GREEN}{rgb}{0,1,0}
 \definecolor{BLUE}{rgb}{0,0,1}
 \definecolor{CYAN}{cmyk}{1,0,0,0}
 \definecolor{MAGENTA}{cmyk}{0,1,0,0}
 \definecolor{YELLOW}{cmyk}{0,0,1,0}
}

\usepackage{color}\usepackage[squaren]{SIunits}

\makeatother

\begin{document}
\flushbottom

\title{Relaxation Oscillations in the Formation of a Polariton Condensate}

\author{Milena De Giorgi}

\affiliation{NNL, Istituto Nanoscienze - Cnr, Via Arnesano, 73100 Lecce, Italy}

\affiliation{CBN-IIT, Istituto Italiano di Tecnologia, Via Barsanti, 73010 Lecce,
Italy}

\author{Dario Ballarini}

\affiliation{CBN-IIT, Istituto Italiano di Tecnologia, Via Barsanti, 73010 Lecce,
Italy}

\affiliation{NNL, Istituto Nanoscienze - Cnr, Via Arnesano, 73100 Lecce, Italy}

\author{Paolo Cazzato}

\affiliation{NNL, Istituto Nanoscienze - Cnr, Via Arnesano, 73100 Lecce, Italy}

\affiliation{CBN-IIT, Istituto Italiano di Tecnologia, Via Barsanti, 73010 Lecce,
Italy}

\author{George Deligeorgis}

\affiliation{CNRS-LAAS and Univ de Toulouse, 7 avenue du colonel Roche, F-31400
Toulouse, France}

\author{Simos I. Tsintzos}

\affiliation{IESL-FORTH, P.O. Box 1527, 71110 Heraklion, Crete, Greece}

\author{Zacharias Hatzopoulos}

\affiliation{IESL-FORTH, P.O. Box 1527, 71110 Heraklion, Crete, Greece}

\affiliation{Department of Physics, University of Crete, 71003 Heraklion, Crete,
Greece}

\author{Pavlos G. Savvidis}

\affiliation{IESL-FORTH, P.O. Box 1527, 71110 Heraklion, Crete, Greece}

\affiliation{Department of Materials Science and Technology, University of Crete,
Greece}

\author{Giuseppe Gigli}

\affiliation{NNL, Istituto Nanoscienze - Cnr, Via Arnesano, 73100 Lecce, Italy}

\affiliation{CBN-IIT, Istituto Italiano di Tecnologia, Via Barsanti, 73010 Lecce,
Italy}

\affiliation{University of Salento, Via Arnesano, 73100 Lecce, Italy}

\author{Fabrice P. Laussy}

\affiliation{Departamento de F\'{i}sica Te\'orica de la Materia Condensada and
Condensed Matter Physics Center (IFIMAC), Universidad Aut\'onoma de
Madrid, 28049 Madrid, Spain}

\author{Daniele Sanvitto}

\affiliation{NNL, Istituto Nanoscienze - Cnr, Via Arnesano, 73100 Lecce, Italy}

\affiliation{CBN-IIT, Istituto Italiano di Tecnologia, Via Barsanti, 73010 Lecce,
Italy}

\begin{abstract}
  We report observation of oscillations in the dynamics of a
  microcavity polariton condensate formed under pulsed non resonant
  excitation.  While oscillations in a condensate have always been
  attributed to Josephson mechanisms due to a chemical potential
  unbalance, here we show that under some localisation conditions of
  the condensate, they may arise from relaxation oscillations, a
  pervasive classical dynamics that repeatedly provokes the sudden
  decay of a reservoir, shutting off relaxation as the reservoir is
  replenished.  Using non-resonant excitation, it is thus possible to
  obtain condensate injection pulses with a record frequency of
  \unit{0.1}{\tera\hertz}.
\end{abstract}


\date{\today}

\maketitle
Light-matter particles---that arise from the strong interaction
between the photonic field in a semiconductor microcavity and the
exciton dipole in quantum wells (QWs)---have recently shown a variety
of interesting phenomena related to non-equilibrium
condensation. These particles, polaritons, have striking
similarities with Bose-Einstein condensates (BECs) of atomic gases,
and, in other aspects, with photon lasers. Nonetheless, polaritons
have demonstrated peculiarities of their own that set them apart from
both quantum condensates and lasers.
Recently many observation in fluid dynamics have shown such
peculiar behaviors, related to the polaritonic dispersion on the one
hand and its dissipative character on the other hand.
Fundamental phenomena like
condensation~\cite{Kasprzak2006,Balili2007},
superfluidity~\cite{Amo2009,Carusotto2004,Amo2008}, quantised
vorticity~\cite{Lagoudakis2008,Sanvitto2009}, quantum
turbulence~\cite{Nardin2011,Sanvitto2011} and phase
transitions~\cite{Cristofolini2013} have shown to deviate from
conventional BEC, stimulating new models for the
quantum dynamics of polariton condensates. These models 
consider some of the many parameters that are typical of polaritons,
including the upper and lower dispersions, pumping, dissipation, spin,
non-linearities, etc. One component that is seldom considered in
details is the exciton reservoir. This, however, often plays an
important role, even under resonant
excitation~\cite{Vishnevsky2012,Wouters2013}, and in particular when
it comes to polariton relaxation and condensation.

\begin{figure}[t]
  \includegraphics[width=0.8\linewidth]{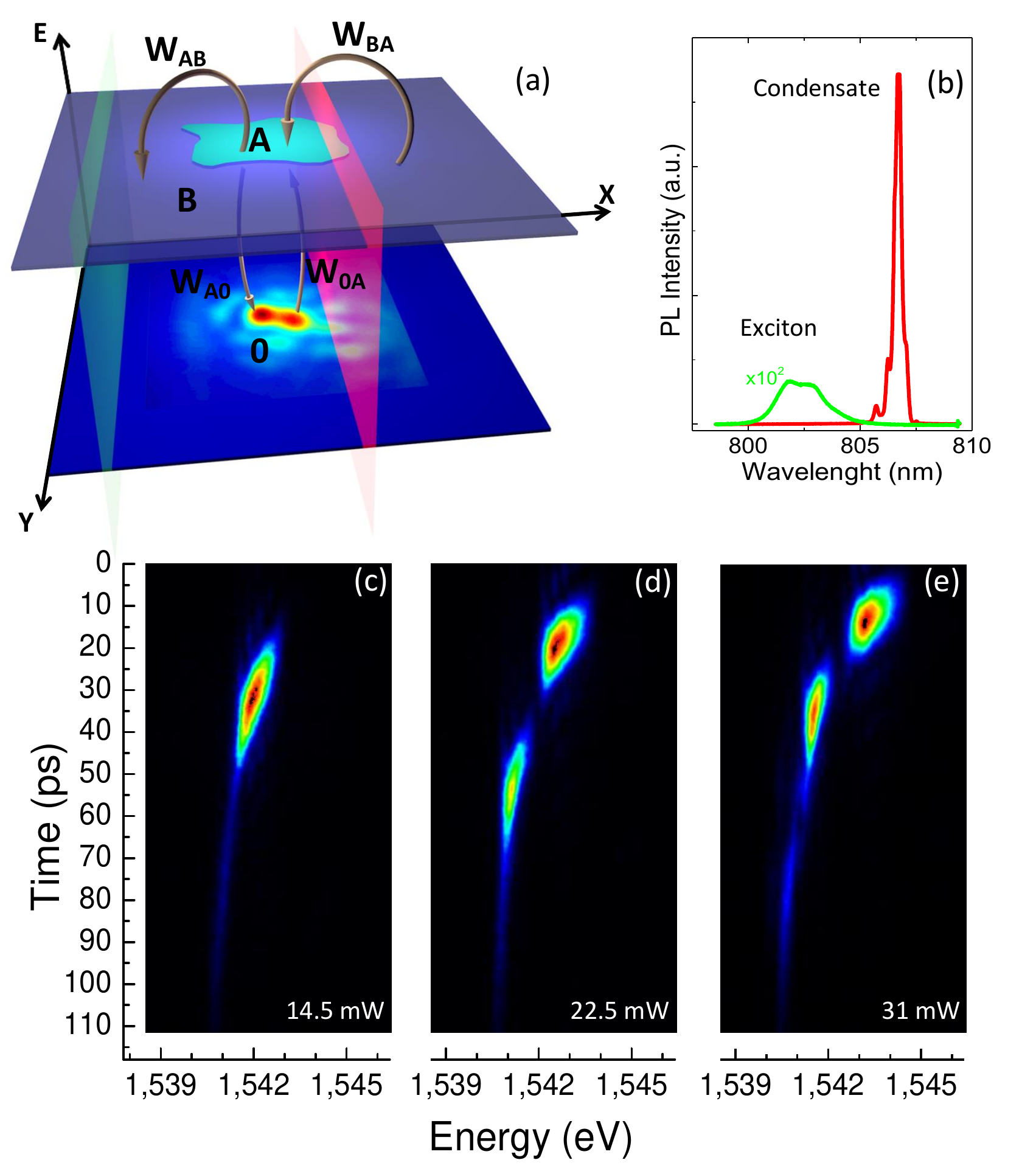} 
  \caption{(Colour online) (a) Sketch of the process of polariton
    condensate formation, considering two kind of exciton reservoirs:
    inside the condensation area (A) and outside (B). (b)
    Photoluminescence collected in the two different regions shown in
    (a) indicated by red and green cross section planes. (c-d) Time
    resolved photoluminescence at different excitation densities
    collected in the same spatial point below (c) and above (d-e)
    $P_{\mathrm{th}}=\unit{15}{\milli\watt}$, the power threshold at
    which the intensity oscillations appear.}
  \label{fig:1} 
\end{figure}

\begin{figure}[t]
  \includegraphics[width=0.8\linewidth]{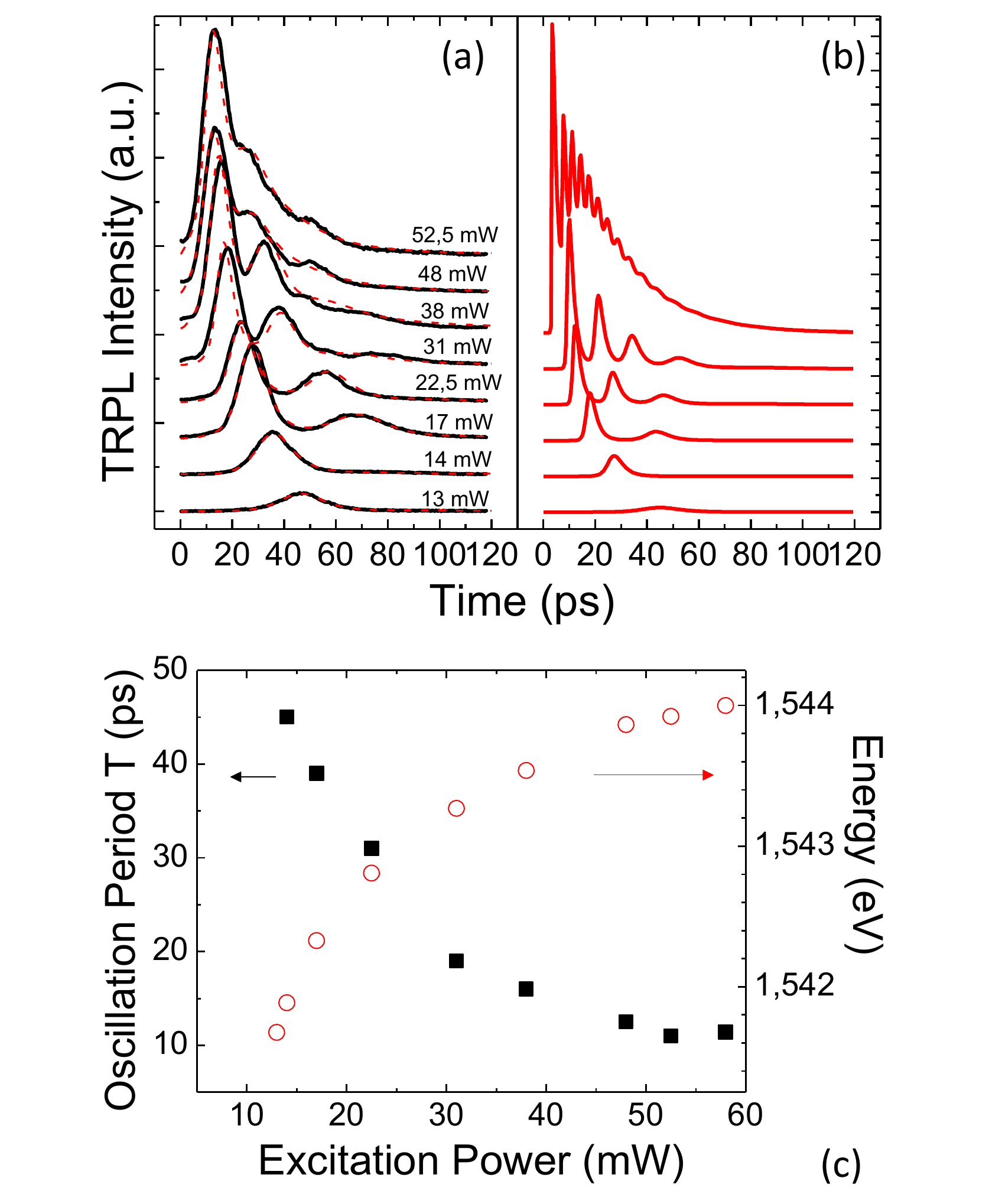} 
  \caption{(Colour online) (a) Relaxation of the condensate as a
    function of time (averaged over the energy) for increasing power
    of the excitation pulse. In dashed are superimposed a theoretical
    fit to the data from semi-classical Boltzmann equations. (b)
    Numerical solutions when the pumping power is the only parameter
    changed: from the bottom up $P_{B}=50,100,200,400,600,5000$.
    {[}Parameters: $\gamma_{0}=0.5\unit{}{\pico\second}^{-1}$,
    $\gamma_{A}=0.01\unit{}{\pico\second}^{-1}$, $\gamma_{B}=0$,
    $\gamma_{P_{B}}=0.05\unit{}{\pico\second}^{-1}$,
    $\gamma_{P_{A}}=0$
    $W_{A\rightarrow0}=0.001\unit{}{\nano\second}^{-1}$,
    $W_{B\rightarrow A}=0.0005\unit{}{\nano\second}^{-1}$,
    $W_{0\rightarrow A}=0$, $W_{A\rightarrow B}=0$, $N_{A}=10$,
    $N_{B}=10^{3}$ , $P_{A}=0${]}.  (c) Period of oscillations and
    energy shift as a function of pumping power.}
  \label{fig:2} 
\end{figure}

We will show here how a polariton condensate formed by non-resonant
excitation, in a confined region of space of a few micron squared, can
exhibit marked oscillations of its population. While many oscillatory
behaviours have been observed~\cite{Lagoudakis2010,Abbarchi2013} or
predicted~\cite{Saito2013} in the polariton literature, relating them
to coherent and/or quantum phenomena, we report semi-classical
oscillations due to the interplay between reservoir feeding and Bose
stimulation. These oscillations fall in the class of so-called
``relaxation oscillations'' \cite{vanderPol1927}, that are well known in class~$B$ lasers
\cite{Siegman1986}, typically solid-state ones with a small active
volume and slow-inversion decay~\cite{vanDruten2000}, where they lead
to the phenomenon of spiking~\cite{Paschotta2008}. The effect is
an important class of self-oscillations~\cite{Jenkins2013}. In this form, it is 
analogous to the Tantalus cup oscillations, used by the Romans to
measure time, or in natural phenomena such as rhythmic springs. The
underlying principle is the following: passed a threshold, a siphon
triggers the rapid emptying of a storage that, once depleted, starts
over a cycle of refilling under continuous pumping. This versatile
physics also takes place in polariton systems in presence of confined
states as sketched in Fig.~1 (a).  In view of
our observations, care should be taken when describing oscillations in
small potential wells, when the total population is not conserved but
constantly fed from an exciton reservoir. In this case, relaxation and
decay cannot be disregarded, being intrinsic to the condensation
process.

The experiments have been carried in an Al$_{0.15}$Ga$_{0.85}$As/AlAs  microcavity with
four sets of 3 GaAs quantum wells (QWs) placed at 
the antinodes of the electric field~\cite{Tosi2012b}.
A pulsed laser of $\unit{100}{\femto\second}$ excites
 non-resonantly the polariton population of the
lower branch (LPB). A 100X objective lens with a numerical aperture of
$\mathrm{NA}=0.7$ is used to collect micro photoluminescence images
resolved in time, space and energy via a streak camera positioned at
the end of a $\unit{550}{\milli\meter}$ spectrometer. We use
excitation spots of \unit{1.2}{\micro\meter} with a spatial resolution
of \unit{0.7}{\micro\meter} in a region of the sample where the amount
of inhomogeneities is high enough to allow for polariton confinement
in areas between 1 to \unit{10}{\micro\meter}$^{2}$ and with confining
potential of \unit{3}{\milli\electronvolt} in
average~\cite{Zajac2012}. Fig.~1 (c-e) shows the dynamics of formation of a
condensate at three different excitation powers. When the excitation
is below $\unit{15}{\milli\watt}$, the condensate emission decays with a
continuous redshift in time~\cite{DelValle2009} caused by a continuous
reduction of the total carriers injected by the laser pulse
into the QWs at time $t_{0}$. At powers above
$P_{\mathrm{th}}=\unit{15}{\milli\watt}$, the decay dynamics changes
abruptly, showing a striking feature that immediately comes across in
Fig.~1(c--e): the formation of the condensate displays an intermittent
emission. This oscillatory behaviour can be followed up to three
periods at the maximum power, with an almost complete suppression of
emission. The time profile of the decays at different excitation
powers is shown Fig.~2(a) (black solid line). From Fig.~1 and
Fig.~2(a), one can see that the number of maxima is connected to the
excitation power and increases for high injection
densities. At low pumping powers, the exciton scattering process which
feeds the condensate occurs only at early times, giving rise to just one 
maximum,  and resulting in a very short lived condensed state 
pulse of a few picoseconds. When the number of
oscillations is sufficiently high, the long decay tail due to the
admixture of exciton recombination and scattering rate is clearly
observed. In Fig.~2(c) the oscillation period (full black dots) is
plotted together with the initial blueshift (empty red dots), which is
proportional to the reservoir population, against the excitation
power. There is a monotonous increase of the oscillation frequency
when the reservoir population is very high, while for low powers, the
oscillations decrease and eventually disappear with low density of the
reservoir. 
Finally, also the rise time depends critically on pumping,
as can however be expected for condensation.
Unlike cavity photons that are non-interacting,
polariton-polariton interactions bring energy into the 
dynamics of relaxation oscillations, resulting in a redshift in time
of the polariton emission.

\begin{figure}[t]
\includegraphics[width=\linewidth]{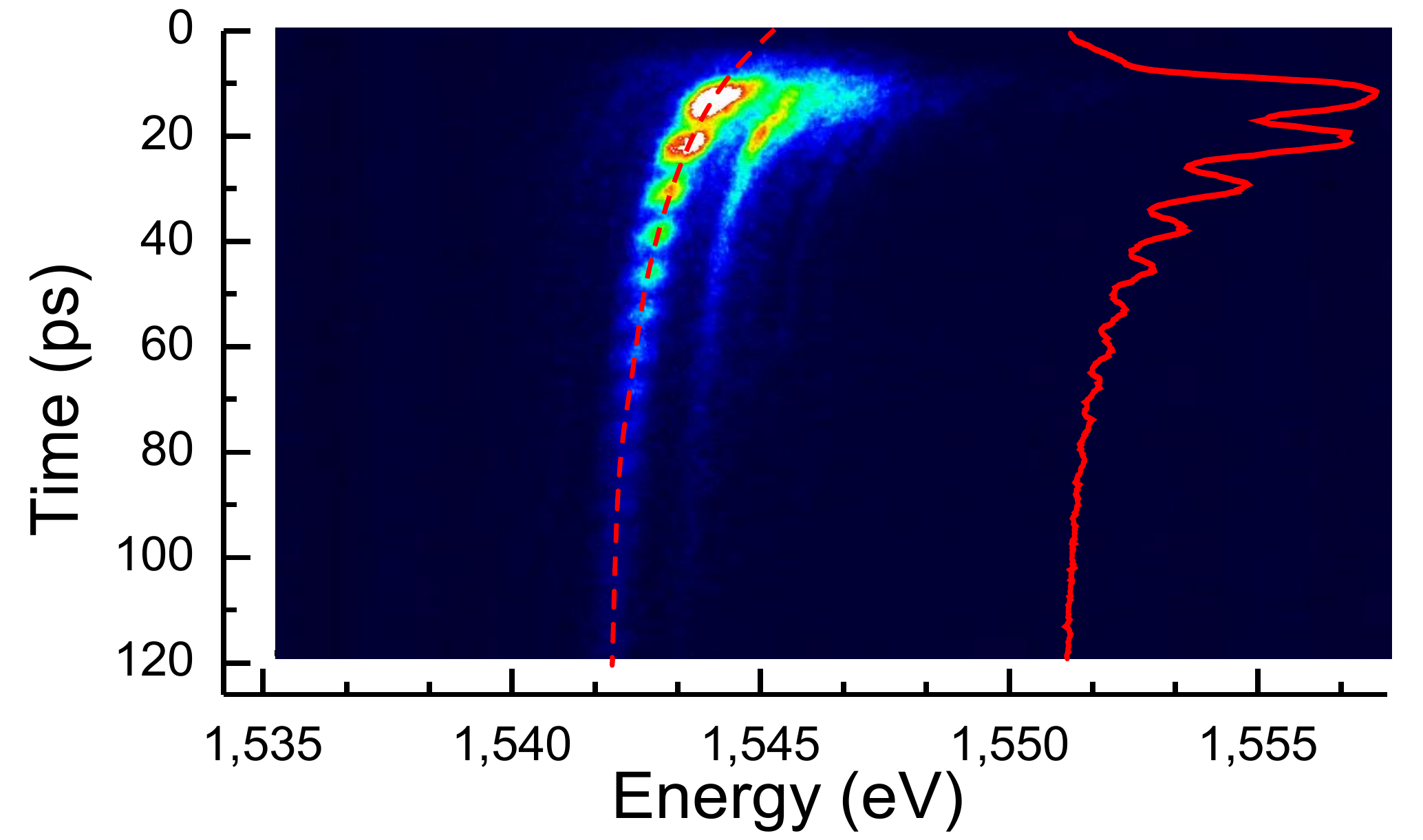} 
\caption{Time resolved photoluminescence at very high power
  $P_{\mathrm{th}}=\unit{100}{\milli\watt}$ collected in a point of
  condensation for slightly bigger confining potential compared to the
  ones in Figs.~1 and~2. The time  profile of the ground state 
is shown by the red curve on the right side of the figure.}
\label{fig:3} 
\end{figure}

By choosing a defect of slightly bigger size, the
oscillation period shrinks to \unit{10}{\pico\second} with
\unit{4}{\pico\second} pulse width, as shown in Fig.~3, allowing for the
observation of more than ten maxima at a record repetition rate of
\unit{0.1}{\tera\hertz}. Relaxation oscillations in polariton condensates could be used
to drive polariton devices~\cite{Ballarini2013} at an extremely fast
clock time, which is still longer than the lifetime-limited switching
time of polariton non-linearities~\cite{Degiorgi2012, Hayat2012}.  Figure~3 also
shows several higher energy trails which belong to excited states of
the same confined well (see also Fig. S2 in the supporting information). 
Given that the energy separation is of only
\unit{1.2}{\milli\electronvolt}, the condensation of a small fraction
of the total population of polaritons is visible for the first, second
and third excited state, as was observed in previous
works~\cite{Sanvitto2009b,Maragkou2010,Galbiati2012}.  This confirms
that the oscillating behaviour is related to the strong confinement
effect. While it mainly affects the ground state, in some cases, we
observe the presence of oscillations also for the first excited
state (Fig. 3).

To show that such oscillations can be explained from the interplay
between the exciton reservoir feeding and Bose stimulation only, we
describe relaxation through semi-classical Boltzmann equations. The
generic equation for the population of the $k$th state reads
$\dot{n}_{k}=-\gamma_{k}n_{k}+P_{k}+(n_{k}+1)\sum_{k'\neq
  k}W_{k'\rightarrow k}n_{k'}-n_{k}\sum_{k'\neq k}W_{k\rightarrow
  k'}(n_{k'}+1)$~\cite{kavokin_book11a}, 
%
%
with $W_{k'\rightarrow k}$ the scattering rate between states $k$
and~$k'$, also with decay rate $\gamma_{k}$ and pumping rate $P_{k}$,
possibly zero. These equations have been extremely successful to
describe the relaxation dynamics of microcavity-polaritons, including
condensation~\cite{tassone1997,porras2002,Kasprzak2008b,Hartwell2010}.
In the case of an infinite system, the quantum number~$k$ is the
wavevector of a plane wave.  In our case, the presence of potentials
makes $k$ a label for unspecified eigenstates, which are the solutions 
of the corresponding Schr\"odinger equation. To reduce the
complexity of the rate equation while retaining the key features of
our system, we separate the set of unspecified $k$ states  into
three groups: the condensate, 0, and two reservoir states, $A$
and~$B$, corresponding to the exciton reservoirs in the localized
condensation spot ($A$) and outside the condensation area ($B$)
[cf.~Fig.~1(a)]. Summing over the states~$k$ with the assumption that
the rates $W_{k\leftrightarrow k'}$ are identical for $k$, $k'$ in any
one of the three groups, and are zero between the condensate and the
upper reservoir ($B$), the polariton populations
$n_{C}=\sum_{k\in\{C\}}n_{k}$ for $C=0,A$ and~$B$ read:
\begin{subequations}
  \label{eq:maroct22105401CEST2013}
  \begin{align}
    \dot{n}_{0}= & W_{A\rightarrow0}(n_{0}+1)n_{A}-W_{0\rightarrow A}n_{0}(n_{A}+N_{A})\nonumber \\
    & -\gamma_{0}n_{0}\,,\\
    \dot{n}_{A}= & W_{0\rightarrow A}(n_{A}+N_{A})n_{0}+W_{B\rightarrow A}(n_{A}+N_{A})n_{B}\nonumber \\
    & -W_{A\rightarrow0}n_{A}(n_{0}+1)-W_{A\rightarrow B}n_{A}(n_{B}+N_{B})\nonumber \\
    & -\gamma_{A}n_{A}+P_{A}\exp(-\gamma_{P_{A}}t)\,,\\
    \dot{n}_{B}= & W_{A\rightarrow B}(n_{B}+N_{B})n_{A}-W_{B\rightarrow A}n_{B}(n_{A}+N_{A})\nonumber \\
    & -\gamma_{B}n_{B}+P_{B}\exp(-\gamma_{P_{B}}t)\,.
  \end{align}
\end{subequations} 

Relaxation oscillations also occur with two classes of states only,
which is the case of the slow-inversion lasers, where the $\beta$
factor---the ratio of photons that go directly into the lasing
mode---determines the oscillatory character~\cite{Note2}. In the polariton case, we
find that the equations above can reproduce fairly well the observed
dynamic within the main numerical constrain of the problem, that is
the one imposed by cavity lifetime,
$\gamma_{0}=\unit{0.5}{\pico\second}^{-1}$, the only parameter known
with high precision. Since our experiment is in the pulsed excitation
regime, we have included a decay rate for the pumping terms,
$\gamma_{P_{A}}$ and $\gamma_{P_{B}}$. In this form, the type of
excitation is left in a general form, describing for instance
diffusion from remote areas. We have also introduced
$N_{C}=\sum_{k\in\{C\}}1$, the number of states in each of the group
$C$ with $C=A,B$. This simple model accounts for the
reservoir and localization as well as drift and diffusion of the excitons.
This is enough to reproduce all the
qualitative features of the experiment, as shown in Fig.~2(a) (red
continuous line), for increasing pumping rate of the higher
reservoir~$P_{B}$.  One finds that, like in the experimental dataset,
the onset of condensation occurs at earlier times with pumping and
with an increasing population in the condensate. Passed a given
pumping intensity, the condensate buildup and decay is followed by a
revival at later times, that also gets closer to the main population
peak with increasing pumping and turns into oscillations.  At high
excitation power, the number of oscillations increases considerably,
as can also be seen in the experimental data (Fig.~3).  While here we
have tuned one parameter only, experimentally, the polariton gas is
extremely complex and most parameters do in fact change with
pumping~\cite{Note3}. Therefore, our model cannot be expected to achieve more than a
qualitative agreement. Nevertheless, allowing more parameters to vary,
an excellent quantitative agreement is obtained already
at the level of approximation of
Eqs.~(\ref{eq:maroct22105401CEST2013}), as shown by the dashed lines
in Fig.~2(a). The oscillations get washed out mainly due to the increasing rates, both
incoming and outgoing, at which the reservoir $A$ is being
(de)populated.  This effective pumping of the reservoir that directly
feeds the condensate can be interpreted as a larger contribution from
an increased number of states whose dynamics becomes independent from
the oscillations of the condensate. When these oscillations are large,
they get imprinted in the reservoir, that reaches a threshold of
stimulated emission leading to its sudden transfer into the ground
state until it empties.  At this point the condensate is not
provisioned anymore and is left to decay, while the reservoir resumes
its slower population growth from higher energy states
reservoir or provided by diffusion, until it
reaches the threshold again, producing another cycle of avalanche into
the ground state. Other sensitive parameters are the rates
$W_{A\leftrightarrow0}$ and $W_{A\leftrightarrow B}$ that increase
with pumping and account for the main patterns as observed in
Fig.~2(b) (to obtain a good agreement, $W_{0\leftrightarrow A}$ and
$W_{A\leftrightarrow B}$ are now nonzero)

\begin{figure}[t]
\includegraphics[width=0.8\linewidth]{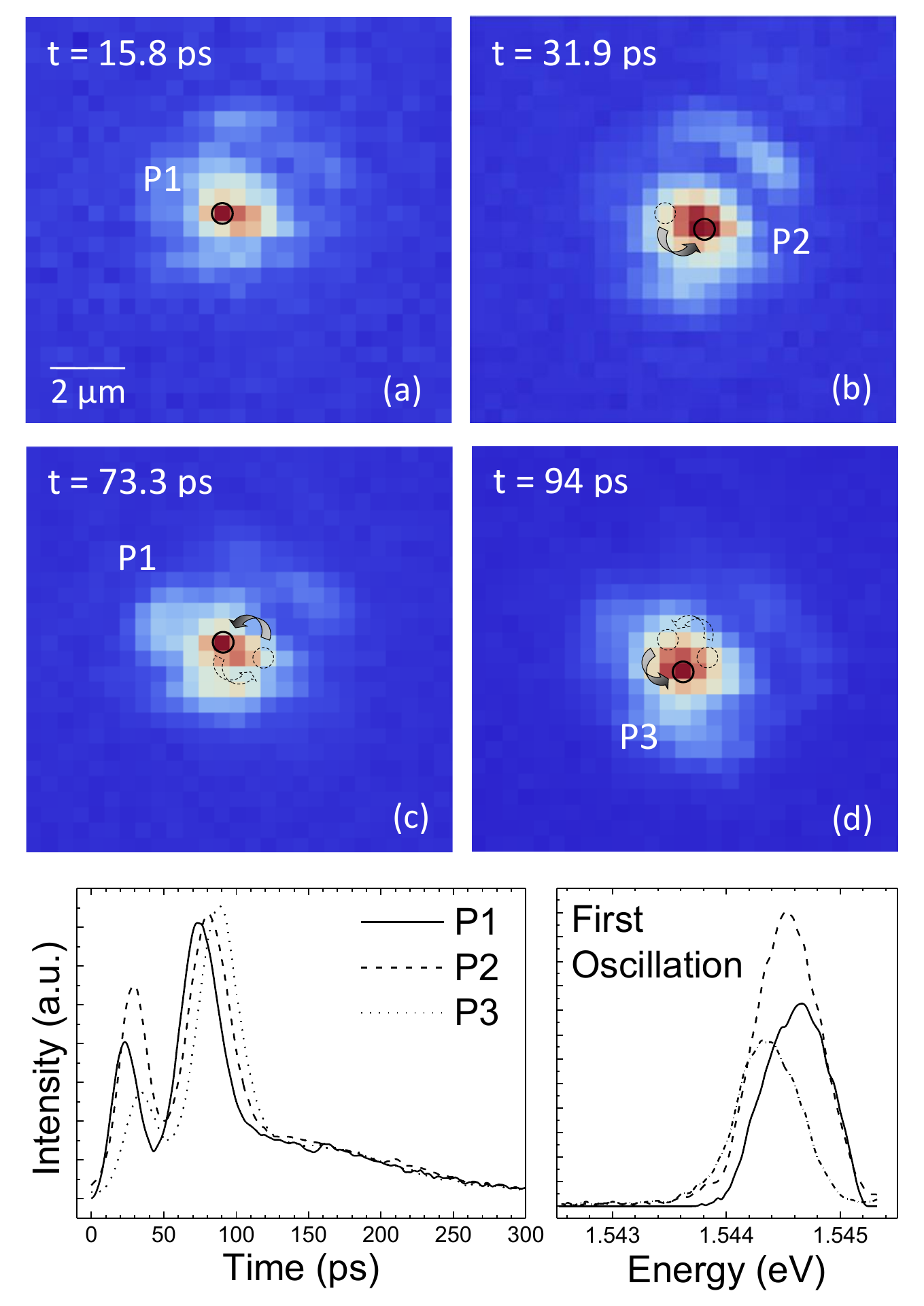} 
\caption{Upper row: Oscillations of the condensate
    in space at different time delays [cf.~supplementary
    video]. Lower row: (a) Time resolved photoluminescence
    intensities extracted from (a-d) in different real space positions
    corresponding to P1-P3. (f) Time integrated spectral profiles of
    the polariton signal matching the oscillation maxima at positions
    P1-P3.}
\label{fig:4} 
\end{figure}

In what follows, we show how the combination of temporal oscillations and
different spatial energy shifts in presence of multiple potential
minima can yield not only relaxation oscillations in time but also be
combined with a spatial movement. This may give the impression of
Josephson oscillations, i.e., spatial oscillations arising from a
small population difference between two superfluid regions separated
by a potential barrier~\cite{Perev1997}. In Fig.~4, we show four
snapshots of a polariton condensate which is forming under similar
conditions as the ones shown in Figs.~1--3, but where condensation
occurs in two areas (separated in space by about one micron) at
different times.  The whole dynamic of the population during
condensation can be better observed in the movie provided in the
supplementary material. Figure~4 shows that the condensate first
appears in position P1 at early times~[Fig.~{4}(a)], then moves to
position P2, which reaches its maximum at time
$t=\unit{32}{\pico\second}$~[Fig.~{4}(b)] and later comes back to
position P1~[Fig.~{4}(c)]. Next it bounces to position
P3~[Fig.~{4}(d)], close to P2, where it reaches again a maximum of
intensity at time $t=\unit{90}{\pico\second}$ and finally comes back
to P1 where it remains till it decays.  Note that for each spatial
point, the maxima of emission shift in time according to the
condensate position at that time frame. This is due to local
difference in the overall redshifts caused by the strong increase and
decrease of the polariton population which undergo condensation. This
is clear by looking at the emission of points P1, P2 and P3 in
Fig.~4({e}) as a function of time. While the condensate is forming,
the emission maximum shifts in time and so does its position in space,
demonstrating a dynamical condensation which is not localised in a
single spot. Indeed, together with an oscillating behaviour, which
depends on the balance between scattering and depletion of the
reservoir, it is also possible to observe a space oscillation which,
instead, depends on energy as shown in Fig.~4{(e-f)}, displaying the
emission intensity from each spatial position of Fig.~{4(a-d)} as a
function of time~({e}) and energy~({f}). In fact, at first the
polariton condensate finds most favourable to form in point P1, which
has the lowest ground state in the spot area. However, as time increases,
the natural redshift---due to a decrease in the total
population---becomes less effective where the higher concentration of
polaritons is present. This allows the formation of the condensate
 in more favourable areas (at P2 in this case). 
Similarly, once the polariton density atP1 has decreased, 
the condensation tends to form back there,
and so on and so forth, giving the impression of synchronized
oscillations. This can be evidenced from the redshift of the first and
second peaks, at each spatial position, shown in Fig.~4(f).
For such inhomogenous condensation area, two
simultaneous effects are observed: on the one hand temporal relaxation
oscillations, due to the strong localisation of the condensate, and on
the other hand, spatial oscillations caused by different blueshift of
the condensation regions.

In conclusions, while the relaxation of polaritons has
been extensively studied, we demonstrate that,
still from the semi-classical standpoint, there
remains rich dynamical behaviours to be explored. We have reported for
the first time in a polariton system an important type of dynamical
instability in a driven dissipative environment, known as relaxation
oscillations,  present in a large class of systems. This
observation is important for future application of polariton as
all-optical devices~\cite{Liew2008, Ballarini2013} and from a
fundamental point of view. They emphasize the crucial role of the
exciton reservoir, that can give rise to nontrivial and striking
dynamical effects.

\begin{acknowledgements}
 Funding from the POLAFLOW ERC Starting
  Grant, Beyond-Nano National Operative Programm PON project,
  Polatom ESF Research Networking Program and "Aristeia" grant
(1978) by Greek GSRT is acknowledged. 
\end{acknowledgements}



\bibliography{Paper_degiorgi}

\end{document}